\documentclass[10pt]{article}
\usepackage{graphicx}
\usepackage{amsmath}
\usepackage{amssymb}
\usepackage{caption2}
\begin{document}
\bibliographystyle{prsty}
\begin{center}
{\large {\bf \sc{  Tetraquark  molecular  states in the $D_s\bar{D}_{s1}$ and $D_s^*\bar{D}_{s0}^*$ mass spectrum  }}} \\[2mm]
Zhi-Gang  Wang \footnote{E-mail: zgwang@aliyun.com.  }, Xiao-Song Yang, Qi Xin     \\
 Department of Physics, North China Electric Power University, Baoding 071003, P. R. China
\end{center}

\begin{abstract}
In the present work, we  construct  the color-singlet-color-singlet type  four-quark  currents  to investigate the  $D_s\bar{D}_{s1}$ and $D_s^*\bar{D}_{s0}^*$ tetraquark molecular states with the $J^{PC}=1^{--}$ and $1^{-+}$ via the QCD sum rules, and obtain satisfactory results.  We can search for the $D_s\bar{D}_{s1}$ and $D_s^*\bar{D}_{s0}^*$ tetraquark molecular states with the $J^{PC}=1^{--}$ and $1^{-+}$ at the BESIII and Belle II in the future.
\end{abstract}



\section{Introduction}

In 2008, the CLEO collaboration measured  the cross sections of the processes   $e^+ e^- \to D_s^+D_s^-$, $D_s^{*+}D_s^-$ and $D_s^{*+}D_s^{*-}$
up to the center-of-mass energy $4.26\,\rm{GeV}$, and observed no evidence of  the $Y(4260)$ \cite{CLEO-DsDs}. In 2010, the BaBar collaboration measured
 the cross sections of the processes   $e^+ e^- \to D_s^+D_s^-$, $D_s^{*+}D_s^-$ and $D_s^{*+}D_s^{*-}$ up to the center-of-mass energy $6.2\,\rm{GeV}$ via initial-state
radiation (ISR), and observed no evidence of  the $Y(4260)$ either \cite{BaBar-DsDs}. Also in 2010, the Belle collaboration measured the  cross sections of the processes   $e^+ e^- \to D_s^+D_s^-$, $D_s^{*+}D_s^-$ and $D_s^{*+}D_s^{*-}$ up to the center-of-mass energy $5.0\,\rm{GeV}$ via initial-state
radiation, and observed that both the  $e^+ e^- \to D_s^{*+}D_s^-$ cross
section and $R$ ratio exhibit an obvious dip near the  mass of the $Y(4260)$ \cite{Belle-DsDs}.
In 2020, the BESIII collaboration measured  the cross sections of the processes $e^+e^-\to D_s^+ D_{s1}(2460)^-$ $+c.c.$ at the center-of-mass energy
$4.467\,\rm{GeV}-4.600\,\rm{GeV}$, and $e^+e^-\to D_s^{\ast +}
D_{s1}(2460)^-$ $+c.c.$ at the center-of-mass energy   $4.590\,\rm{GeV}-4.600\,\rm{GeV}$, and observed no obvious charmonium or charmonium-like structure \cite{BESIII-DsDs-2020}.  Recently, the BESIII collaboration measured the  cross sections of the processes $e^+e^-\to D_s^{*+}D_{s0}^*(2317)^- +c.c.$ and $e^+e^-\to
D_s^{*+}D_{s1}(2460)^- +c.c.$  at the center-of-mass energy $ 4.600\,\rm{GeV} - 4.700\,\rm{GeV}$, and $e^+e^-\to D_s^{*+}D_{s1}(2536)^- +c.c.
$  at the center-of-mass energy $ 4.660\,\rm{GeV} - 4.700\,\rm{GeV}$, and observed  no structure in either process  \cite{BESIII-DsDs-2021}.

The assignments of the $Y(4260)$, such as the tetraquark state
\cite{Y4260-tetraquark-Maiani,WZG-Y4260-1,WZG-Y4260-2,WZG-Y4260-3},
tetraquark molecular state \cite{Y4260-molecule}, hybrid state
\cite{Y4260-Hybrid-1,Y4260-Hybrid-2}, conventional charmonium
\cite{Y4260-cc-1,Y4260-cc-2},  are still in hot debate. If there
exist the color-singlet-color-singlet type tetraquark  states
$c\bar{s}s\bar{c}$, irrespective of weak bound states or higher
resonances, they can decay to their constituents, the $c\bar{s}$
and $s\bar{c}$ color-singlet clusters, through the
Okubo-Zweig-Iizuka super-allowed fall-apart mechanism saving
feasible in the phase-space. When the experimental data are
accumulated, the BESIII and Belle II collaborations maybe  observe
them in the $D_s\bar{D}_s$, $D^*_s\bar{D}_s$, $D^*_s\bar{D}^*_s$,
$D_s\bar{D}_{s1}$, $D_s^*\bar{D}_{s1}$, $D^*_s\bar{D}_{s0}^*$,
$\cdots$ invariant mass spectrum in the $e^+e^-$ scattering
processes, which can shed light on the nature of the $X$, $Y$ and
$Z$ states. It is necessary and important to investigate the mass
spectrum of the $D_s\bar{D}_s$, $D_s\bar{D}_s^*$,
$D_s^*\bar{D}_s^*$, $D_s\bar{D}_{s1}$, $D_s^*\bar{D}_{s1}$,
$D_s^{*}\bar{D}_{s0}^*$, $\cdots$ tetraquark molecular states and
make reliable predictions.

In Ref.\cite{WZG-DD-IJMPA},  we accomplish  the operator product
expansion for the correlation functions up to the vacuum condensates
of dimension $10$ consistently  and investigate  the  ground state hidden-charm
tetraquark molecular states  without strange, with strange and with hidden-strange,
such as the $D\bar{D}^*$, $D^*\bar{D}^*$, $D\bar{D}_s^*$,
$D_s\bar{D}_s^*$,  $D^*\bar{D}_s^*$ and $D_s^*\bar{D}_s^*$ tetraquark molecular
 states with the $J^{PC}=0^{++}$, $1^{++}$, $1^{+-}$ and $2^{++}$, via the QCD sum rules
 comprehensively, and make possible assignments of the existing $X$, $Y$ and $Z$ states,
 such as the $X_c(3872)$, $Z_c(3900)$, $Z_{cs}(3985/4000)$, $Z_c(4020/4025)$.

In Ref.\cite{WZG-DD-Vector}, we construct   the color-singlet-color-singlet type
four-quark currents   to explore   tetraquark molecular states $D\bar{D}_1(2420)$
and $D^*\bar{D}_0^*(2400)$ with the $J^{PC}=1^{--}$ and $1^{-+}$ via the QCD sum rules
by calculating the contributions of the vacuum condensates up to dimension-10. The
predictions only support assigning the $Y(4390)$  to be the $D\bar{D}_1$ molecular
state with the $J^{PC}=1^{--}$.

 In Ref.\cite{WZG-Landau}, we construct the color-singlet-color-singlet type
 tensor current  to explore  the neutral $D_s^*\bar{D}_{s1}-D_{s1}\bar{D}_s^*$
 tetraquark molecular states with the $J^{PC}=1^{-+}$ via the QCD sum rules, and obtain
 the molecule mass  $4.67\pm0.08\,\rm{GeV}$, which is in very good agreement
 with the mass of the $X(4630)$ observed later by the LHCb  collaboration,
 $M_{X(4630)}=4626 \pm 16 {}^{+18}_{-110}\,\rm{MeV}$ \cite{LHCb-X4685}.

All in all, the QCD sum rules is a powerful theoretical approach in exploring
the masses and decay widths of the $X$, $Y$ and $Z$ states, and
has achieved many successful descriptions  in the scenario of
tetraquark states
\cite{Narison-3872,Azizi-Z4430-tetra-6,WZG-bb-cc-tetraquark-1,WZG-bb-cc-tetraquark-2,
SLZhu-tetra-AV-8,HXChen-tetra-8,CFQiao-Zcs3985-tetra-mol-8,
WangHuangtao-2014-PRD,Wang-tetra-formula-1,Wang-tetra-formula-2}
or tetraquark molecular states
\cite{JRZhang-mole-6,JRZhang-Z10610-mol-6,Nielsen-X3872-mol-cc-8,Narison-Z4430-mol-8,WZG-mole-formula-1,WZG-mole-formula-2}.

In the present work, we extend our previous works  to explore  tetraquark molecular states $D_s\bar{D}_{s1}$ and $D_s^*\bar{D}_{s0}^*$ with the $J^{PC}=1^{--}$ and $1^{-+}$ via the
QCD sum rules by accomplishing the operator product expansion up to the vacuum condensates of dimension 10 consistently, and make possible predictions to be confronted to the experimental data at the BESIII and Belle II in the future.

The article is arranged as follows:  we obtain the QCD sum rules for the  vector tetraquark  molecular states  in section 2; in section 3, we present the numerical results and discussions; section 4 is reserved for our conclusion.

\section{QCD sum rules for  the  vector tetraquark molecular states }
Let us write down  the  correlation functions $\Pi_{\mu\nu}(p)$  in the QCD sum rules,
\begin{eqnarray}
\Pi_{\mu\nu}(p)&=&i\int d^4x e^{ip \cdot x} \langle0|T\left\{J_\mu(x)J_\nu^{\dagger}(0)\right\}|0\rangle \, ,
\end{eqnarray}
where the color-singlet-color-singlet type four-quark currents $J_\mu(x)=J_\mu^1(x),\,J_\mu^2(x),\,J_\mu^3(x),\,J_\mu^4(x)$,
\begin{eqnarray}
J^1_\mu(x)&=&\frac{1}{\sqrt{2}}\Big\{ \bar{s}(x)i\gamma_5c(x)\bar{c}(x)\gamma_\mu \gamma_5 s(x)-\bar{s}(x)\gamma_\mu \gamma_5c(x)\bar{c}(x)i\gamma_5 s(x)\Big\} \, , \nonumber \\
J^2_\mu(x)&=&\frac{1}{\sqrt{2}}\Big\{ \bar{s}(x)i\gamma_5c(x)\bar{c}(x)\gamma_\mu \gamma_5 s(x)+\bar{s}(x)\gamma_\mu \gamma_5c(x)\bar{c}(x)i\gamma_5 s(x)\Big\} \, , \nonumber \\
J^3_\mu(x)&=&\frac{1}{\sqrt{2}}\Big\{ \bar{s}(x)c(x)\bar{c}(x)\gamma_\mu  s(x)+\bar{s}(x)\gamma_\mu c(x)\bar{c}(x) s(x)\Big\} \, , \nonumber \\
J^4_\mu(x)&=&\frac{1}{\sqrt{2}}\Big\{ \bar{s}(x)c(x)\bar{c}(x)\gamma_\mu  d(x)-\bar{s}(x)\gamma_\mu c(x)\bar{c}(x) s(x)\Big\} \, .
\end{eqnarray}
 Under charge conjugation transform $\widehat{C}$, the currents $J_\mu(x)$ have the properties,
\begin{eqnarray}
\widehat{C}J^{1/3}_{\mu}(x)\widehat{C}^{-1}&=& - J^{1/3}_\mu(x)  \, , \nonumber\\
\widehat{C}J^{2/4}_{\mu}(x)\widehat{C}^{-1}&=&+ J^{2/4}_\mu(x) \, ,
\end{eqnarray}
the currents $J_\mu(x)$ are  eigenstates of the charge conjugation.

In the present work, we choose the local
color-singlet-color-singlet type four-quark currents to
interpolate the hidden-charm tetraquark states, which have two
color-singlet clusters. The color-singlet clusters have the same
quantum numbers as the charmed mesons, such as $D_s$, $D_s^*$,
$D^*_{s0}$ and $D_{s1}$, except for the masses, as those
color-singlet clusters are not necessary to be the physical
mesons.

The physical mesons with two valence quarks  are spatial extended
objects and have mean spatial sizes  about the magnitude $\sqrt{\langle r^2\rangle}\sim 0.5\,\rm{fm}$. In the present work or in the QCD sum rules,
 though we refer the color-singlet-color-singlet
type tetraquark states as the tetraquark molecular states, they
are not the usually called molecular states. They  have the mean/average spatial sizes as that of the typical heavy mesons, and are compact
objects.  The usually called molecular states are loosely bound
 states consist of the physical mesons, the mean spatial sizes are proportional
 to the inverse of the binding  energies,  about $1\,\rm{fm}$ or larger than $1\,\rm{fm}$, which are too large to be interpolated by
 the local four-quark currents.

 The currents $\bar{s}(x)i\gamma_5 c(x)$ and $\bar{s}(x)\gamma_\mu c(x)$ have the
spin-parity $J^P=0^-$ and $1^-$, respectively, and couple potentially to the mesons $D_s$ and $D_s^*$, respectively.  While
the currents $\bar{s}(x)i\gamma_5 \gamma_5c(x)$ and $\bar{s}(x)\gamma_\mu \gamma_5 c(x)$ have the spin-parity
$J^P=0^+$ and $1^+$, respectively, and couple potentially to the mesons  $D^*_{s0}$ and $D_{s1}$, respectively, as multiplying
$\gamma_5$ to the currents changes their parity, the net effects of the relative P-wave are embodied implicitly in the $\gamma_5$.
In the heavy quark limit, the total angular momentum of a
heavy-light meson $\vec{J}=\vec{j}_\ell+\vec{S}_Q$, the light quark total angular
momentum $\vec{j}_\ell=\vec{L}+\vec{S}_{\bar{q}}$, the $\vec{L}$
is the light quark angular momentum,  the $\vec{S}_Q$ and
$\vec{S}_{\bar{q}}$ are the heavy quark and light quark spins,
respectively. There exist  two doublets $(0^+,1^+)$ and
$(1^+,2^+)$ for $j_{\ell}=\frac{1}{2}$ and $\frac{3}{2}$,
respectively. On the other hand, the heavy-light mesons  can also
be classified in terms of eigenvalues of the light quark angular
momentum $\left|^{2S+1}L_J \right> $, $\vec{J}=\vec{S}+\vec{L}$,
$\vec{S}=\vec{S}_{\bar{q}}+\vec{S}_Q$ is a sum of the intrinsic
quark spins.

In the present case, the two doublets are
$(D_{s0}^*(2317),D_{s1}^\prime(2460))$ and
$(D_{s1}(2536),D_{s2}^*(2573))$. We usually choose the currents
$\bar{s}(x)c(x)$ and $\bar{s}(x)\gamma_\mu \gamma_5 c(x)$, which
have the total angular momenta, $\vec{J}=\vec{S}+\vec{L}$, i.e.
$\vec{0}=\vec{1}+\vec{1}$ and $\vec{1}=\vec{1}+\vec{1}$
respectively, to interpolate the $D_{s0}^*(2317)$ and
$D_{s1}^\prime(2460)$ respectively \cite{WZG-decay-EPJC}. In fact,
there exists mixing effect between the $j_{\ell}=\frac{1}{2}$ and
$\frac{3}{2}$ states with the spin-parity $J^P=1^+$,
\begin{eqnarray}
\left( {\begin{array}{*{20}{c}}
   {\left| {{1^ + },j_\ell = {{\frac{1}
{2}} }} \right\rangle }  \\
   {\left| {{1^ + },j_\ell = {{\frac{3}
{2}} }} \right\rangle }  \\
 \end{array} }\right)  = \left( {\begin{array}{*{20}{c}}
   {\cos \theta } & { - \sin \theta }  \\
   {\sin \theta } & {\cos \theta }  \\
 \end{array} } \right)
\left(  {\begin{array}{*{20}{c}}
   {\left| {^3{P_1}} \right\rangle }  \\
   {\left| {^1{P_1}} \right\rangle }  \\
 \end{array} }\right) \, ,
\end{eqnarray}
where the mixing angle $\tan \theta=\frac{1}{\sqrt{2}}$ in the heavy quark limit
\cite{Matsuki-mix}. So we cannot exclude the coupling between the
current $\bar{s}(x)\gamma_\mu \gamma_5 c(x)$ and meson
$D_{s1}(2536)$. In the present work, we will not distinguish the
$D_{s1}^\prime(2460)$ and $D_{s1}(2536)$, and use the notation
$D_{s1}$ to represent the color-singlet cluster with the
spin-parity  $J^P=1^+$.

At the hadron side of the correlation functions $\Pi_{\mu\nu}(p)$, we  isolate the contributions of the ground state  vector tetraquark molecular  states $Y$,
\begin{eqnarray}
\Pi_{\mu\nu}(p)&=&\frac{\lambda_{Y}^2}{M_{Y}^2-p^2}\left(-g_{\mu\nu} +\frac{p_\mu p_\nu}{p^2}\right) +\cdots \, \, , \nonumber\\
               &=&\Pi(p^2)\left(-g_{\mu\nu} +\frac{p_\mu p_\nu}{p^2}\right) +\cdots \, \, ,
\end{eqnarray}
where the pole residues  $\lambda_{Y}$ are defined by $ \langle 0|J_\mu(0)|Y(p)\rangle=\lambda_{Y} \,\varepsilon_\mu$,
the $\varepsilon_\mu$ are the polarization vectors.

 We accomplish  the operator product expansion up to the vacuum condensates  of dimension-10 consistently and assume vacuum saturation for the  higher dimensional  vacuum condensates,   and write the correlation functions $\Pi(p^2)$ at the QCD side in the form,
 \begin{eqnarray}
 \Pi(p^2)&=&\frac{1}{\pi}\int_{4m_c^2}^{\infty} ds \frac{{\rm Im}\Pi(s)}{s-p^2}\, ,
 \end{eqnarray}
   through dispersion relation.
In calculations, we contract the  $s$ and $c$ quark fields in the
correlation functions  with Wick theorem, and substitute the quark
lines with the full $s$ and $c$ quark propagators $S_{ij}(x)$ and
$C_{ij}(x)$, respectively,
 \begin{eqnarray}\label{L-quark}
S_{ij}(x)&=& \frac{i\delta_{ij}\!\not\!{x}}{ 2\pi^2x^4}
-\frac{\delta_{ij}m_s}{4\pi^2x^2}-\frac{\delta_{ij}\langle
\bar{s}s\rangle}{12} +\frac{i\delta_{ij}\!\not\!{x}m_s
\langle\bar{s}s\rangle}{48}-\frac{\delta_{ij}x^2\langle
\bar{s}g_s\sigma Gs\rangle}{192}+\frac{i\delta_{ij}x^2\!\not\!{x}
m_s\langle \bar{s}g_s\sigma
 Gs\rangle }{1152}\nonumber\\
&& -\frac{ig_s G^{a}_{\alpha\beta}t^a_{ij}(\!\not\!{x}
\sigma^{\alpha\beta}+\sigma^{\alpha\beta}
\!\not\!{x})}{32\pi^2x^2}
-\frac{1}{8}\langle\bar{s}_j\sigma^{\mu\nu}s_i \rangle
\sigma_{\mu\nu}+\cdots \, ,
\end{eqnarray}

\begin{eqnarray}\label{H-quark}
C_{ij}(x)&=&\frac{i}{(2\pi)^4}\int d^4k e^{-ik \cdot x} \left\{
\frac{\delta_{ij}}{\!\not\!{k}-m_c}
-\frac{g_sG^n_{\alpha\beta}t^n_{ij}}{4}\frac{\sigma^{\alpha\beta}(\!\not\!{k}+m_c)+(\!\not\!{k}+m_c)
\sigma^{\alpha\beta}}{(k^2-m_c^2)^2}\right.\nonumber\\
&&\left. -\frac{g_s^2 (t^at^b)_{ij} G^a_{\alpha\beta}G^b_{\mu\nu}(f^{\alpha\beta\mu\nu}+f^{\alpha\mu\beta\nu}+f^{\alpha\mu\nu\beta}) }{4(k^2-m_c^2)^5}+\cdots\right\} \, ,\nonumber\\
f^{\alpha\beta\mu\nu}&=&(\!\not\!{k}+m_c)\gamma^\alpha(\!\not\!{k}+m_c)\gamma^\beta(\!\not\!{k}+m_c)\gamma^\mu(\!\not\!{k}+m_c)\gamma^\nu(\!\not\!{k}+m_c)\,
,
\end{eqnarray}
 to facilitate the cumbersome tasks,
where  $t^n=\frac{\lambda^n}{2}$,  the $\lambda^n$ is the
Gell-Mann matrix \cite{Reinders85}. In the full $s$-quark
propagator, see Eq.\eqref{L-quark}, the $s$-quark mass $m_s$ is
taken as a mall quantity and is treated perturbatively, direct
calculations indicate that such a perturbative treatment of the
$s$-quark mass does not modify the dispersion relation comparing
to the massless light quarks. For more technical details, one can
consult
Refs.\cite{WangHuangtao-2014-PRD,Wang-tetra-formula-1,Wang-tetra-formula-2,WZG-mole-formula-1,WZG-mole-formula-2}.

   Then  we  implement the
quark-hadron duality below the continuum thresholds $s_0$ and accomplish Borel transform  in regard to
the variable $P^2=-p^2$ to acquire  the  QCD sum rules,
\begin{eqnarray}\label{QCDSR}
\lambda^2_{Y}\, \exp\left(-\frac{M^2_{Y}}{T^2}\right)= \int_{4m_c^2}^{s_0} ds\, \rho_{QCD}(s) \, \exp\left(-\frac{s}{T^2}\right) \, ,
\end{eqnarray}
the explicit expressions of the QCD spectral densities $\rho_{QCD}$ are available via contacting the corresponding author by E-mail.

 In the present work, we take  account of the vacuum condensates $\langle\bar{s}s\rangle$, $\langle\frac{\alpha_{s}GG}{\pi}\rangle$, $\langle\bar{s}g_{s}\sigma Gs\rangle$, $\langle\bar{s}s\rangle^2$, $g_s^2\langle\bar{s}s\rangle^2$,
$\langle\bar{s}s\rangle \langle\frac{\alpha_{s}GG}{\pi}\rangle$,
$\langle\bar{s}s\rangle  \langle\bar{s}g_{s}\sigma Gs\rangle$,
$\langle\bar{s}g_{s}\sigma Gs\rangle^2$ and
$\langle\bar{s}s\rangle^2 \langle\frac{\alpha_{s}GG}{\pi}\rangle$.
The four-quark condensate $g_s^2\langle \bar{s}s\rangle^2$
originates from the matrix elements $\langle \bar{s}\gamma_\mu t^a
s g_s D_\eta G^a_{\lambda\tau}\rangle$,
$\langle\bar{s}_jD^{\dagger}_{\mu}D^{\dagger}_{\nu}D^{\dagger}_{\alpha}s_i\rangle$
and $\langle\bar{s}_jD_{\mu}D_{\nu}D_{\alpha}s_i\rangle$, rather than
originates from the radiative corrections of the $\langle
\bar{s}s\rangle^2$, the strong fine structure constant
$\alpha_s=\frac{g_s^2}{4\pi}$ appears at the tree level. We adopt
the truncations $n\leq 10$ and $k\leq 1$  consistently, the
operators of the orders $\mathcal{O}( \alpha_s^{k})$ with $k> 1$
are  discarded
\cite{WangHuangtao-2014-PRD,Wang-tetra-formula-1,Wang-tetra-formula-2,WZG-mole-formula-1,WZG-mole-formula-2}.
 The condensates $\langle g_s^3 GGG\rangle$, $\langle \frac{\alpha_s GG}{\pi}\rangle^2$,
 $\langle \frac{\alpha_s GG}{\pi}\rangle\langle \bar{s} g_s \sigma Gs\rangle$ have the dimensions of mass 6, 8, 9 respectively,  but they are  vacuum expectations
of the operators of the order    $\mathcal{O}( \alpha_s^{3/2})$, $\mathcal{O}(\alpha_s^2)$, $\mathcal{O}( \alpha_s^{3/2})$ respectively, and are discarded for a consistent treatment.

 We  differentiate  Eq.\eqref{QCDSR} in regard  to  $\tau=\frac{1}{T^2}$, and eliminate the
 pole residues  $\lambda_{Y}$ to acquire the QCD sum rules for the molecule masses,
 \begin{eqnarray}
 M^2_{Y}= \frac{-\frac{d}{d \tau }\int_{4m_c^2}^{s_0} ds\rho_{QCD}(s)e^{-\tau s}}{\int_{4m_c^2}^{s_0} ds \rho_{QCD}(s)e^{-\tau s}}\, .
\end{eqnarray}

\section{Numerical results and discussions}
At the QCD side, we adopt  the standard values of the  vacuum condensates
$\langle\bar{q}q \rangle=-(0.24\pm 0.01\, \rm{GeV})^3$,  $\langle\bar{s}s \rangle=(0.8\pm0.1)\langle\bar{q}q \rangle$,
  $\langle\bar{s}g_s\sigma G s \rangle=m_0^2\langle \bar{s}s \rangle$,
$m_0^2=(0.8 \pm 0.1)\,\rm{GeV}^2$, $\langle \frac{\alpha_s
GG}{\pi}\rangle=(0.33\,\rm{GeV})^4$    at the energy scale
$\mu=1\, \rm{GeV}$
\cite{Reinders85,SVZ79-1,SVZ79-2,Colangelo-Review}, and  take the
$\overline{MS}$ masses $m_{c}(m_c)=(1.275\pm0.025)\,\rm{GeV}$
 and $m_s(\mu=2\,\rm{GeV})=(0.095\pm0.005)\,\rm{GeV}$
 from the Particle Data Group \cite{PDG}.
In addition,  we take account of
the energy-scale dependence of  the quark condensates, mixed quark condensates and $\overline{MS}$ masses in regard  to  the renormalization group equation \cite{Narison-mix},
 \begin{eqnarray}
  \langle\bar{s}s \rangle(\mu)&=&\langle\bar{s}s \rangle({\rm 1 GeV})\left[\frac{\alpha_{s}({\rm 1 GeV})}{\alpha_{s}(\mu)}\right]^{\frac{12}{33-2n_f}}\, , \nonumber\\
   \langle\bar{s}g_s \sigma Gs \rangle(\mu)&=&\langle\bar{s}g_s \sigma Gs \rangle({\rm 1 GeV})\left[\frac{\alpha_{s}({\rm 1 GeV})}{\alpha_{s}(\mu)}\right]^{\frac{2}{33-2n_f}}\, ,\nonumber\\
m_c(\mu)&=&m_c(m_c)\left[\frac{\alpha_{s}(\mu)}{\alpha_{s}(m_c)}\right]^{\frac{12}{33-2n_f}} \, ,\nonumber\\
m_s(\mu)&=&m_s({\rm 2GeV} )\left[\frac{\alpha_{s}(\mu)}{\alpha_{s}({\rm 2GeV})}\right]^{\frac{12}{33-2n_f}}\, ,\nonumber\\
\alpha_s(\mu)&=&\frac{1}{b_0t}\left[1-\frac{b_1}{b_0^2}\frac{\log t}{t} +\frac{b_1^2(\log^2{t}-\log{t}-1)+b_0b_2}{b_0^4t^2}\right]\, ,
\end{eqnarray}
 as the strong fine-structure constant $\alpha_s$ already appears at the tree level,  where $t=\log \frac{\mu^2}{\Lambda^2}$, $b_0=\frac{33-2n_f}{12\pi}$, $b_1=\frac{153-19n_f}{24\pi^2}$, $b_2=\frac{2857-\frac{5033}{9}n_f+\frac{325}{27}n_f^2}{128\pi^3}$,  $\Lambda=213\,\rm{MeV}$, $296\,\rm{MeV}$  and  $339\,\rm{MeV}$ for the quark flavors  $n_f=5$, $4$ and $3$, respectively  \cite{PDG}.
As we investigate  the  tetraquark molecular states with hidden-charm and hidden-strange, it is naturel  to  choose the quark flavors $n_f=4$, and evolve the QCD spectral densities $\rho_{QCD}(s)$ to the suitable   energy scales $\mu$ to extract the molecule masses.

In the present work, we  take the energy scale formula
$\mu=\sqrt{M^2_{X/Y/Z}-(2{\mathbb{M}}_c)^2}$ with the updated
value of the effective $c$-quark mass
${\mathbb{M}}_c=1.85\,\rm{GeV}$ to acquire the suitable energy
scales of the QCD spectral densities
\cite{WZG-DD-Vector,WZG-mole-formula-1,WZG-mole-formula-2}. We
introduce the effective heavy quark mass $\mathbb{M}_c$ and divide
the tetraquark molecular states into both the heavy degrees of
freedom $2{\mathbb{M}}_c$ and light  degrees of freedom
$\mu=\sqrt{M^2_{X/Y/Z/T}-(2{\mathbb{M}}_c)^2}$  by
setting $m_u=m_d=0$. We can also consider the light flavor $SU(3)$
breaking effects, and acquire  the light degrees of freedom
$\mu=\sqrt{M^2_{X/Y/Z/T}-(2{\mathbb{M}}_c)^2}-2\,m_s(\mu)$, in
other words,
$\mu+2\,m_s(\mu)=\sqrt{M^2_{X/Y/Z/T}-(2{\mathbb{M}}_c)^2}$.

We can rewrite the energy scale formula in the form,
\begin{eqnarray}\label{formula-Regge}
M^2_{X/Y/Z}&=&\mu^2+{\rm Constants}\, ,
\end{eqnarray}
where the Constants have the value $4{\mathbb{M}}_c^2$ and fitted
by the QCD sum rules
\cite{Wang-tetra-formula-1,Wang-tetra-formula-2,WZG-mole-formula-1,WZG-mole-formula-2},
the predicted tetraquark (molecule) masses and the
pertinent/suitable  energy scales of the QCD spectral densities
have a  Regge-trajectory-like relation \cite{WZG-formular-Regge}.
In calculations, we take account of the light-flavor $SU(3)$
mass-breaking effects by subtracting a small $s$-quark mass to
obtain the modified energy scale formula
$\mu=\sqrt{M^2_{X/Y/Z}-(2{\mathbb{M}}_c)^2}-2m_s(\mu)$. Analysis
of the $J/\psi$ and $\Upsilon$ with the famous  Cornell potential,
i.e. the Coulomb-potential-plus-linear-potential, leads to the
constituent quark masses $m_c=1.84\,\rm{GeV}$ and
$m_b=5.17\,\rm{GeV}$ \cite{Cornell}. We can  set the effective
$c$-quark mass ${\mathbb{M}}_c=m_c=1.84\,\rm{GeV}$, which is
consistent with the   updated value
${\mathbb{M}}_c=1.85\,\rm{GeV}$. In numerical calculations, we add
an uncertainty $\delta\mu=\pm 0.1\,\rm{GeV}$ considering the
uncertainty of the ${\mathbb{M}}_c$.

At the beginning, we tentatively set the masses of the molecular
states to be the sum of the physical masses of the two charmed
mesons, which correspond to the two color-singlet clusters inside
the molecular states, and obtain the energy scales $\mu$ through
the modified energy scale formula. Then we calculate the molecule
masses with the QCD sum rules by searching for  the best Borel parameters $T^2$ and continuum threshold parameters $s_0$ via
trial and error, and examine whether or not the
modified energy scale formula is satisfied. We vary the molecule
masses therefore the energy scales $\mu$ slowly and steadily until
reach the satisfactory results, and acquire the  Borel parameters, continuum threshold parameters,
pole contributions and optimal energy scales,  which are shown plainly in Table \ref{Borel}.

From Table \ref{Borel}, we can
see that the central values of the pole contributions are larger than $50\%$,
the pole dominance criterion can be  satisfied very well. In calculations, we observe that in the Borel windows, the dominant contributions come from the
perturbative terms, the  contributions come from the vacuum condensates  of dimension $10$, the highest dimensional vacuum condensates,   are much less than $1\%$, the convergent behaviors of the operator product expansion are very good.

\begin{figure}
\centering
\includegraphics[totalheight=6cm,width=7cm]{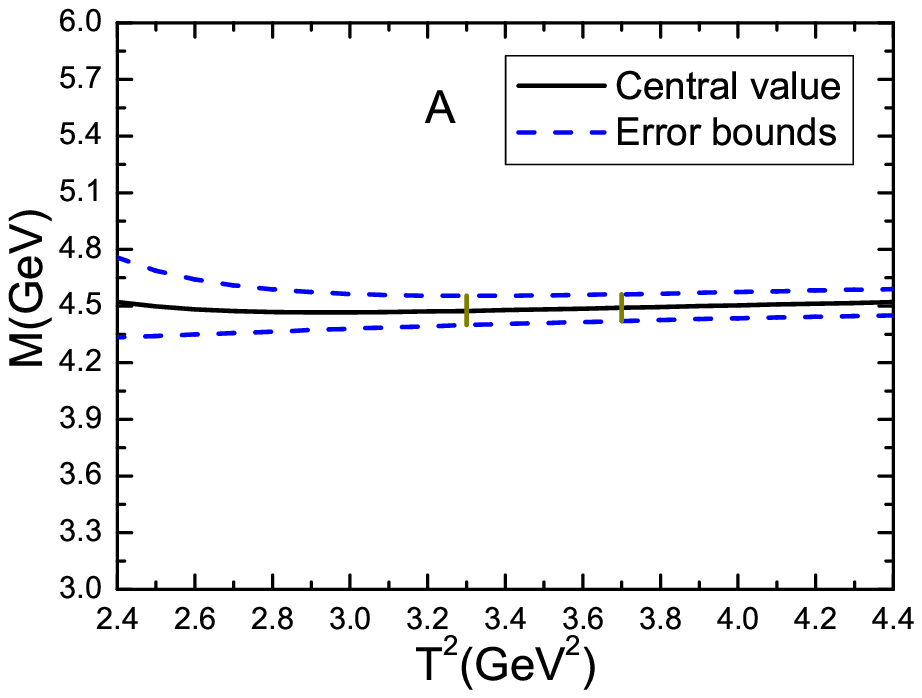}
\includegraphics[totalheight=6cm,width=7cm]{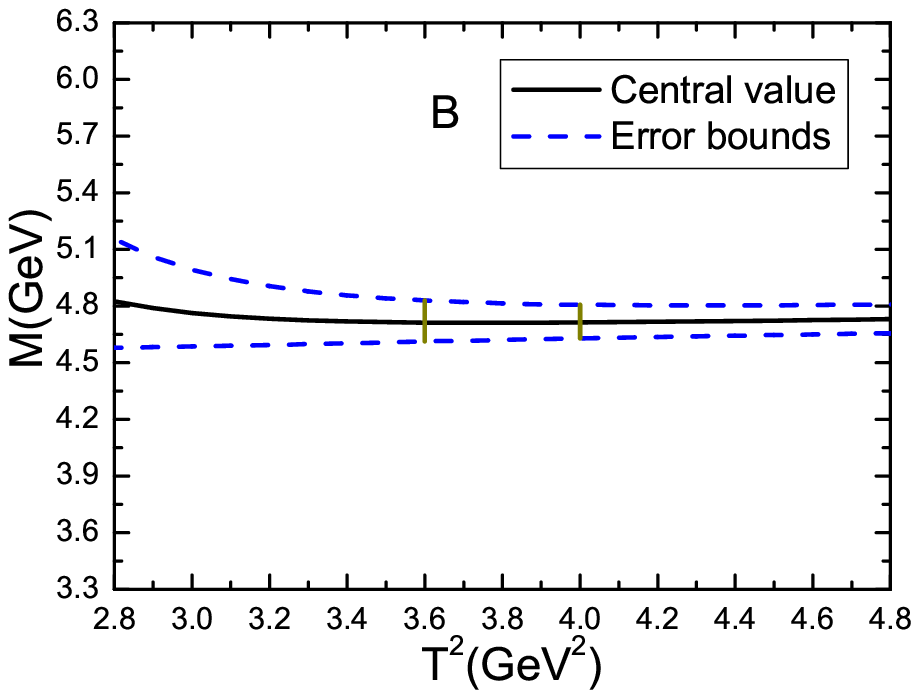}
\includegraphics[totalheight=6cm,width=7cm]{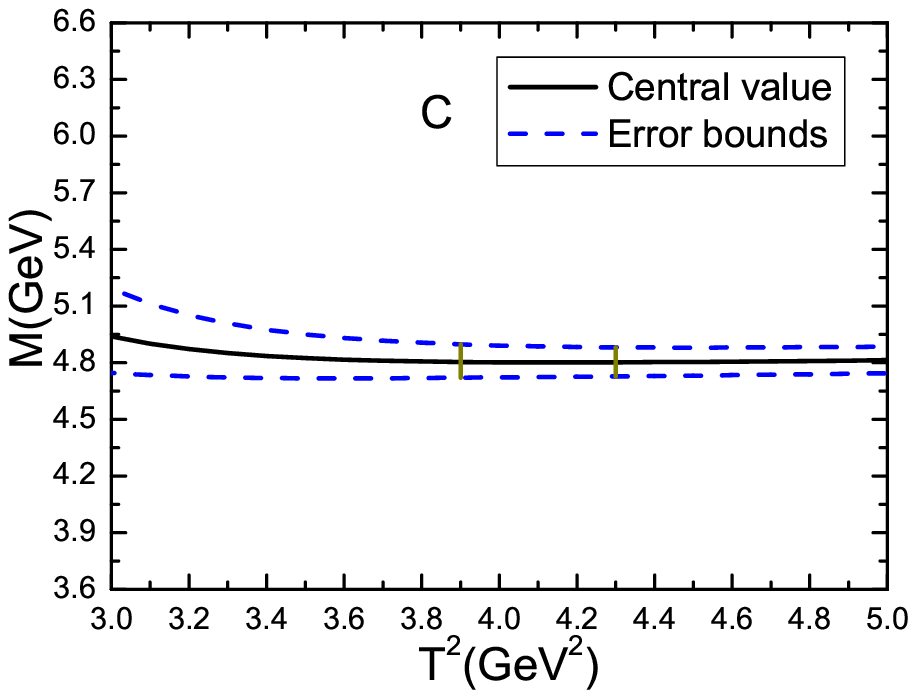}
\includegraphics[totalheight=6cm,width=7cm]{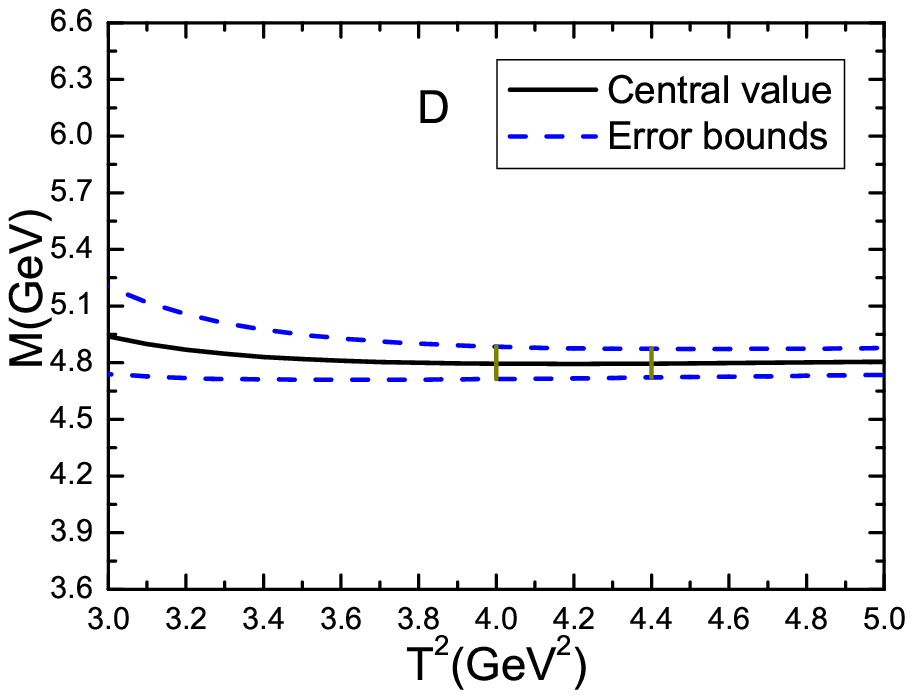}
  \caption{ The masses  with variations of the  Borel parameters $T^2$, where  the $A$, $B$, $C$ and $D$ correspond to the tetraquark  molecular states
  $D_s\bar{D}_{s1}(1^{--})$, $D_s\bar{D}_{s1}(1^{-+})$, $D_s^*\bar{D}_{s0}^*(1^{--})$ and $D_s^*\bar{D}_{s0}^*(1^{-+})$, respectively. The regions between the two short vertical lines are the Borel windows.   }\label{mass-Borel}
\end{figure}

Now we take account of all uncertainties of the input parameters,
and obtain the values of the masses and pole residues of
 the   vector tetraquark molecular states with hidden-charm and hidden-strange, which are  shown plainly in Fig.\ref{mass-Borel} or Table 1. From Table \ref{Borel}, we can infer that  the threshold parameters and the predicted masses  satisfy the relation $\sqrt{s_0}=M_{Y}+(0.4\sim 0.6)\,\rm{GeV}$, which is consistent with our naive expectation, in addition, we can also infer that the modified energy scale formula is well satisfied.
 In Fig.\ref{mass-Borel},  we plot the tetraquark molecule masses  with variations
of the Borel parameters at much larger intervals   than the  Borel windows, where the regions between the two short vertical lines are the Borel windows. From the figure, we can see clearly that there appear platforms  really  in the Borel windows, and we expect to make reasonable predictions, which can be confronted to the experimental data in the future.

From Table \ref{Borel}, we can see that the mass-splitting for the
two molecular states with the $J^{PC}=1^{--}$ is rather large, while the mass-splitting for the two molecular states with the
$J^{PC}=1^{-+}$ is rather  small. We simplify the analysis to explore the origination of the mass-splittings by taking the limit $m_s\to
0$. For the $D_s\bar{D}_{s1}$ and $D_s^*\bar{D}_{s0}^*$ molecular states with the $J^{PC}=1^{--}$, the contributions of the quark
condensate $\langle\bar{s}{s}\rangle$ are zero,   the contributions of the mixed condensate
$\langle\bar{s}g_s \sigma  G{s}\rangle$   are of the same
magnitude but opposite sign. For the $D_s\bar{D}_{s1}$ and
$D_s^*\bar{D}_{s0}^*$ molecular states with the $J^{PC}=1^{-+}$,
the contributions of the vacuum condensates
$\langle\bar{s}{s}\rangle$ and $\langle\bar{s}g_s \sigma
G{s}\rangle$ are canceled out  with each other severely, the net
contributions are very small. The different contributions of the
vacuum condensates $\langle\bar{s}{s}\rangle$ and
$\langle\bar{s}g_s \sigma G{s}\rangle$ lead to the different
behaviors of the mass-splittings of the molecular states with the $J^{PC}=1^{--}$ and $1^{-+}$.

In Table \ref{assignment}, we also  present the predictions for  the masses of the hidden-charm tetraquark molecular states with one
P-wave constituent (or color-singlet cluster) in our previous works \cite{WZG-DD-Vector,WZG-Landau,Di-ZY}. The predictions
$M_{D\bar{D}_1(1^{--})}=4.36\pm0.08\,\rm{GeV}$ and $M_{D_s^*\bar{D}_{s1}(1^{-+})}=4.67\pm0.08\,\rm{GeV}$ are
consistent with the experimental data $M_{Y(4390)}=4391.6\pm6.3\pm1.0\,\rm{MeV}$ and $M_{X(4630)}=4626
\pm 16 {}^{+18}_{-110}\,\rm{MeV}$ respectively within
uncertainties \cite{LHCb-X4685,BES-Y4390}, and support assigning
the $Y(4390)$ and $X(4630)$   to be the $D\bar{D}_1(1^{--})$ and
$D_s^*\bar{D}_{s1}(1^{-+})$ tetraquark  molecular states,
respectively.

The $X(4630)$ was observed in the $J/\psi \phi$ invariant mass
spectrum in the $B^+\to J/\psi \phi K^+$ decays  by the LHCb
collaboration and has the quantum numbers $J^{PC}=1^{-+}$
\cite{LHCb-X4685}. The $Y(4630)$ was observed in the
$\Lambda_c^+\Lambda_c^-$ invariant mass spectrum  in the exclusive
process $e^+e^- \to \gamma_{ISR} \Lambda_c^+ \Lambda_c^-$ by the
Belle collaboration, and has  the mass
$M_Y=4634^{+8}_{-7}{}^{+5}_{-8} \,\rm{MeV}$ and width
$\Gamma_Y=92^{+40}_{-24}{}^{+10}_{-21}\,\rm{MeV}$, respectively
\cite{Belle-LamLam-4630}, which  are consistent  with that of the
charmonium-like state $Y(4660)$ within errors \cite{PDG}. However,
precise measurement of the  $e^+e^-\to\Lambda_c^+\Lambda_c^-$
cross section near the threshold  by the BESIII collaboration
 indicates  that there maybe exist a bound state below the
$\Lambda_c^+\Lambda_c^-$ threshold, which differs from the
$Y(4630/4660)$ remarkably \cite{Belle-Lamlam,GuoFK-LamLam}. In
Ref.\cite{WZG-Y4660-decay}, we take the $Y(4660)$ as the
tetraquark state with the $J^{PC}=1^{--}$ and study the strong
decays $Y(4660)\to J/\psi f_0(980)$, $ \eta_c \phi$, $
\chi_{c0}\phi$, $ D_s \bar{D}_s$, $ D_s^* \bar{D}^*_s$, $ D_s
\bar{D}^*_s$,  $ D_s^* \bar{D}_s$, $ \psi^\prime \pi^+\pi^-$,
$J/\psi\phi$ with the QCD sum rules based on  rigorous
quark-hadron quality, and  observe that the decay to $J/\psi \phi$
is greatly suppressed or forbidden. The predicted width
$\Gamma(Y(4660) )= 74.2^{+29.2}_{-19.2}\,{\rm{MeV}}$ supports
assigning the $Y(4660)$ to be the
$[sc]_P[\bar{s}\bar{c}]_A-[sc]_A[\bar{s}\bar{c}]_P$ type
tetraquark state with the $J^{PC}=1^{--}$.  The $X(4630)$,
$\Lambda_c^+\Lambda_c^-$ resonance and $Y(4660)$ are different
particles.

At the present time, there are no experimental  candidates for the
$D_s\bar{D}_{s1}$ and $D_s^*\bar{D}_{s0}^*$ tetraquark molecular
states with the $J^{PC}=1^{--}$ and $1^{-+}$.  In the  scenario of
tetraquark molecular states, we can assign the $X(3872)$,
$Z_c(3900/3885)$,   $Z_{cs}(3985/4000)$, $Z_c(4020/4025)$,
$Y_c(4390)$, $X_c(4630)$ and $Z_b(10610/10650)$ to be the
tetraquark molecular states tentatively based on the predicted
masses from our previous QCD sum rules calculations
\cite{WZG-DD-IJMPA,WZG-DD-Vector,WZG-Landau,WZG-mole-formula-1,WZG-mole-formula-2},
\begin{eqnarray}
X(3872)&=&\frac{1}{\sqrt{2}}\left( D\overline{D}^{*} -  D^{*}\overline{D}\right) \,\,\,({\rm with}\,\,\,1^{++})\, , \nonumber \\
Z_c(3900/3885)&=&\frac{1}{\sqrt{2}}\left( D\overline{D}^{*} +  D^{*}\overline{D}\right)\,\,\,({\rm with}\,\,\,1^{+-}) \, , \nonumber\\
Z_c(4020/4025)&=& D^*\overline{D}^{*} \,\,\,({\rm with}\,\,\,1^{+-})  \, ,\nonumber \\
Z_{cs}(3985/4000)&=&\frac{1}{\sqrt{2}}\left( D\overline{D}_s^{*} \mp  D^{*}\overline{D}_s\right) \,\,\,({\rm with}\,\,\, 1^{+\pm})\, , \nonumber\\
Y_{c}(4390)&=&\frac{1}{\sqrt{2}}\left( D\bar{D}_{1}-D_{1}\bar{D}\right) \,\,\,({\rm with}\,\,\, 1^{--})\, , \nonumber\\
X_{c}(4630)&=&\frac{1}{\sqrt{2}}\left( D_s^*\bar{D}_{s1}-D_{s1}\bar{D}_s^*\right) \,\,\,({\rm with}\,\,\, 1^{-+})\, , \nonumber\\
Z_b(10610)&=&\frac{1}{\sqrt{2}}\left( B\overline{B}^* + B^*\overline{B}\right)\,\,\,({\rm with}\,\,\,1^{+-}) \, ,\nonumber \\
Z_b(10650)&=&  B^{*}\overline{B}^{*} \,\,\,({\rm with}\,\,\,1^{+-}) \, ,
\end{eqnarray}
the $D_s\bar{D}_{s1}$ molecular states were also discussed in
Ref.\cite{WangFL} recently.

\begin{table}
\begin{center}
\begin{tabular}{|c|c|c|c|c|c|c|c|}\hline\hline
                            &$T^2(\rm{GeV}^2)$ &$\sqrt{s_0}(\rm{GeV})$ &pole        &$\mu(\rm{GeV})$  &$M_{Y}(\rm{GeV})$ &$\lambda_{Y}(10^{-2}\rm{GeV}^5)$\\ \hline

$D_s\bar{D}_{s1}$ ($1^{--}$)     &$3.3-3.7$         &$5.0\pm0.1$            &$(45-64)\%$ &$2.3$            &$4.48\pm0.08$     &$4.47\pm 0.61 $    \\ \hline

$D_s\bar{D}_{s1}$ ($1^{-+}$)     &$3.6-4.0$         &$5.2\pm0.1$            &$(44-63)\%$ &$2.7$            &$4.71\pm 0.10$    &$5.89 \pm0.74$     \\ \hline

$D_s^*\bar{D}_{s0}^*$ ($1^{--}$) &$3.9-4.3$         &$5.3\pm0.1$            &$(44-62)\%$ &$2.9$            &$4.80\pm0.08$     &$7.46\pm 0.87$    \\ \hline

$D_s^*\bar{D}_{s0}^*$ ($1^{-+}$) &$4.0-4.4$         &$5.3\pm0.1$            &$(43-61)\%$ &$2.9$            &$4.79\pm0.08$     &$7.49\pm 0.86$    \\ \hline \hline
\end{tabular}
\end{center}
\caption{ The Borel parameters, continuum threshold parameters, pole contributions, energy scales, masses and pole residues of the vector tetraquark  molecular  states. }\label{Borel}
\end{table}

\begin{table}
\begin{center}
\begin{tabular}{|c|c|c|c|c|c|c|c|}\hline\hline
       Molecule                        &$M_{Y}(\rm{GeV})$  &Assignment        &  References          \\ \hline

$D\bar{D}_1$ ($1^{--}$)                &$4.36\pm0.08$      & ? $Y(4390)$      &  \cite{WZG-DD-Vector}  \\ \hline

$D^*\bar{D}_0^*$ ($1^{--}$)            &$4.78\pm0.07$      &                  &  \cite{WZG-DD-Vector}  \\ \hline

$D\bar{D}_1$ ($1^{-+}$)                &$4.60\pm 0.08$     &                  &  \cite{WZG-DD-Vector}  \\ \hline

$D^*\bar{D}_0^*$ ($1^{-+}$)            &$4.73\pm0.07$      &                  &  \cite{WZG-DD-Vector} \\ \hline

$D_s\bar{D}_{s1}$ ($1^{--}$)           &$4.48\pm0.08$      &                  &  This work       \\ \hline

$D_s^*\bar{D}_{s0}^*$ ($1^{--}$)       &$4.80\pm0.08$      &                  &  This work   \\ \hline

$D_s\bar{D}_{s1}$ ($1^{-+}$)           &$4.71\pm 0.10$     &                  &  This work  \\ \hline

$D_s^*\bar{D}_{s0}^*$ ($1^{-+}$)       &$4.79\pm0.08$      &                  &  This work  \\ \hline

$D_s^*\bar{D}_{s1}$ ($1^{-+}$)         &$4.67\pm0.08$      &  ? $X(4630)$     &  \cite{WZG-Landau}  \\ \hline

$D\bar{D}_{s0}^*$ ($0^{-}$)            &$4.61\pm0.10$      &                  &  \cite{Di-ZY}  \\ \hline

$D^*\bar{D}_{s1}$ ($0^{-}$)            &$4.60\pm0.07$      &                  &  \cite{Di-ZY}  \\ \hline

\hline
\end{tabular}
\end{center}
\caption{ Assignments of the tetraquark molecular states with one P-wave cluster. }\label{assignment}
\end{table}

Now let us perform Fierz re-arrangement for the four-quark  currents $J_\mu$  both in the color space and Dirac-spinor  space,
\begin{eqnarray}
2\sqrt{2}J_\mu^1&=&\frac{1}{3} i\bar{s}\gamma_\mu s\, \bar{c}c -\frac{1}{3}i\bar{s} s\, \bar{c}\gamma_\mu c -\frac{1}{3}\bar{s}\gamma^\beta\gamma_5 s\, \bar{c}\sigma_{\mu\beta}\gamma_5 c +\frac{1}{3}\bar{s}\sigma_{\mu\beta} \gamma_5s\, \bar{c}\gamma^\beta\gamma_5 c+\cdots \, , \nonumber\\
2\sqrt{2}J_\mu^2&=& \frac{1}{3}\bar{s}\sigma_{\mu\beta} s\, \bar{c}\gamma^\beta c+\frac{1}{3}\bar{s}\gamma^\beta s\, \bar{c}\sigma_{\mu\beta} c-\frac{1}{3}\bar{s}i\gamma_5 s\, \bar{c}\gamma_\mu\gamma_5c -\frac{1}{3}\bar{s}\gamma_\mu\gamma_5 s\, \bar{c}i\gamma_5 c+\cdots \, , \nonumber\\
2\sqrt{2}J_\mu^3&=& -\frac{1}{3}\bar{s}\gamma_\mu s\, \bar{c}c -\frac{1}{3}\bar{s} s\, \bar{c}\gamma_\mu c -\frac{1}{3}i\bar{s}\gamma^\beta\gamma_5 s\, \bar{c}\sigma_{\mu\beta}\gamma_5 c -\frac{1}{3}i\bar{s}\sigma_{\mu\beta} \gamma_5s\, \bar{c}\gamma^\beta\gamma_5 c+\cdots \, , \nonumber\\
2\sqrt{2}J_\mu^4&=& -\frac{1}{3}i\bar{s}\sigma_{\mu\beta} s\, \bar{c}\gamma^\beta c+\frac{1}{3}i\bar{s}\gamma^\beta s\, \bar{c}\sigma_{\mu\beta} c+\frac{1}{3}i\bar{s}i\gamma_5 s\, \bar{c}\gamma_\mu\gamma_5c -\frac{1}{3}i\bar{s}\gamma_\mu\gamma_5 s\, \bar{c}i\gamma_5 c+\cdots \, ,
\end{eqnarray}
to illustrate the two-body strong decays.
The components $\bar{s}\Gamma s\, \bar{c}\Gamma^\prime c$  with $\Gamma, \Gamma^\prime=1,\,\gamma_\mu,\, \gamma_\mu\gamma_5,\, \cdots$ couple potentially to a series of $c\bar{c}$-$s\bar{s}$-type meson-pairs, or $c\bar{c}s\bar{s}$-type tetraquark  molecular states, which decay to their components via the Okubo-Zweig-Iizuka super-allowed fall-apart mechanism.
We can investigate the  $D_s\bar{D}_{s1}$ and $D_s^*\bar{D}_{s0}^*$ tetraquark molecular states with the $J^{PC}=1^{--}$ and $1^{-+}$ through the two-body strong decays,
\begin{eqnarray}
  D_s\bar{D}_{s1}(1^{--}) &\to&  \chi_{c0}\phi \, ,\, J/\psi f_0(980) \, ,\,  J/\psi f_1(1285)\, ,\, J/\psi \eta\, ,\, h_c \eta \, ,\, \eta_c\phi\, ,\, \chi_{c1}\phi\, , \,\eta_c h_1(1415)\, , \nonumber\\
  D_s\bar{D}_{s1}(1^{-+}) &\to& J/\psi\phi \, ,\, J/\psi h_1(1415)\, ,\,h_c\phi \, ,\, \chi_{c1} \eta \, ,\,  \eta_c \eta\, ,\,  \eta_c f_1(1285)\, , \nonumber\\
 D_s^*\bar{D}_{s0}^*(1^{--}) &\to&  \chi_{c0}\phi \, ,\, J/\psi f_0(980) \, ,\,  J/\psi f_1(1285)\, ,\, J/\psi \eta\, ,\, h_c \eta \, ,\, \eta_c\phi\, ,\, \chi_{c1}\phi\, , \,\eta_c h_1(1415)\, , \nonumber\\
 D_s^*\bar{D}_{s0}^*(1^{-+}) &\to&  J/\psi\phi \, ,\, J/\psi h_1(1415)\, ,\,h_c\phi \, ,\, \chi_{c1} \eta \, ,\,  \eta_c \eta\, ,\,  \eta_c f_1(1285)\, ,
\end{eqnarray}
besides the final states $D_s\bar{D}_{s1}$, $D_{s1}\bar{D}_{s}$,  $D_s^*\bar{D}_{s0}^*$ and $D_{s0}^*\bar{D}_{s}^*$.

\section{Conclusion}
In this article, we  construct   the color-singlet-color-singlet type four-quark  currents   to investigate the  $D_s\bar{D}_{s1}$ and $D_s^*\bar{D}_{s0}^*$ tetraquark molecular states with the $J^{PC}=1^{--}$ and $1^{-+}$ via the QCD sum rules.  We accomplish the operator product expansion up to  the vacuum condensates of dimension-10  consistently, take account of the $SU(3)$ mass-breaking effects and adopt the modified energy scale formula to choose the best  energy scales of the QCD spectral densities, then extract the masses and pole residues of the tetraquark molecular states in the suitable Borel windows, which satisfy the two fundamental criteria of the QCD sum rules. We can search for the $D_s\bar{D}_{s1}$ and $D_s^*\bar{D}_{s0}^*$ tetraquark molecular states with the $J^{PC}=1^{--}$ and $1^{-+}$ at the BESIII and Belle II in the future, and confront the present predictions to the experimental data.

\section*{Acknowledgements}
This  work is supported by National Natural Science Foundation, Grant Number  11775079.

\end{document}